\documentclass[conference]{IEEEtran}
\IEEEoverridecommandlockouts
\usepackage{cite}
\usepackage{amsmath,amssymb,amsfonts}
\usepackage{algorithmic}
\usepackage{graphicx}
\usepackage{textcomp}
\usepackage{xcolor} 
\usepackage{amsmath,amssymb,amsfonts}
   
\usepackage{multicol}
\usepackage{multirow} 
\usepackage{url} 
\usepackage{hyperref} 
\usepackage{moreverb}
\usepackage{colortbl}
\usepackage{diagbox} 
 \usepackage{balance}

\begin{document}

\title{Monitoring Informed Testing for IoT 
\thanks{PREPRINT.}
}

 
 \author{\IEEEauthorblockN{Ahmed Abdullah}
\IEEEauthorblockA{\textit{School of Science} \\
\textit{RMIT University}\\
Melbourne, Australia\\
ahmed.abdullah@rmit.edu.au}
\and
\IEEEauthorblockN{Heinz W. Schmidt}
\IEEEauthorblockA{\textit{School of Science} \\
\textit{RMIT University}\\
Melbourne, Australia\\
heinrich.schmidt@rmit.edu.au}
\and
\IEEEauthorblockN{Maria Spichkova}
\IEEEauthorblockA{\textit{School of Science} \\
\textit{RMIT University}\\
Melbourne, Australia\\
maria.spichkova@rmit.edu.au}
\and
\IEEEauthorblockN{Huai Liu}
\IEEEauthorblockA{\textit{Engineering \& Science} \\
\textit{Victoria University}\\
Melbourne, Australia\\
huai.liu@vu.edu.au} 
}

\maketitle

\begin{abstract}
Internet of Things (IoT) systems continuously collect a large amount of data from heterogeneous ``smart objects'' through standardised service interfaces. 
A key challenge is how to use these data and relevant event logs to construct continuously adapted usage profiles and apply them to enhance testing methods, 
i.e., prioritization of tests for the testing of continuous integration of an IoT system. In addition, these usage profiles provide relevance weightings to analyse architecture and behaviour of the system. Based on the analysis, testing methods can predict specific system locations that are susceptible to error, and therefore suggest where expanded runtime monitoring is necessary.    
Furthermore, IoT aims to connect billions of ``smart devices'' over the network. 
Testing even a small IoT system connecting a few dozens of smart devices would require 
a network of test Virtual Machines (VMs) possibly spreading across the Fog and the Cloud.  
In this paper we propose a framework for testing of each IoT layer in a separate VM environment, and discuss potential difficulties with optimal VM allocation.
\end{abstract}


\section{Introduction}
\label{sec:introduction}
 
Internet of Things (IoT) aims to connect billions of ``smart devices'' over the internet. 
Usefulness of IoT systems is being realised in various application domains such as transportation, healthcare, smart cities, smart cars, smart factories, etc. Depending on the complexity of an IoT system, its architecture may involve several layers of networking, where appropriate IoT middleware support an IoT application. 
One of common middleware design approaches is Service Oriented Architecture (SOA), cf. \cite{razzaque2016middleware}.
SOA-based IoT middleware is a promising design perspective to overcome heterogeneity of physical devices and their underlying communication technologies with high level standardised service interfaces.

IoT network and middleware design follow a common layered approach. 
Both from networking and middleware perspective, Fog and Cloud computing are envisioned as natural fit for layered design of IoT systems, cf. \cite{yannuzzi2014key,bonomi2012fog}. 
Layered architecture model of IoT middleware allow designing IoT application components, 
which reside into separate VMs that may spread across Cloud and Fog infrastructure. 
Cloud infrastructure provides centralized resources and processing at a global scale \cite{yusuf2015chiminey,spichkova2016managing,spichkova2015scalable,spichkova2016towards}. As Cloud computing and storage services are highly scalable and efficient, many applications utilise Cloud for analytics, Big Data processing, etc. 
It also enables long running and data-intensive scientific experiments \cite{yusuf2017chiminey}. 
In the Fog infrastructure, certain services are managed at the edge of the network, at so-called \emph{Fog nodes}. 
Both Fog and Cloud nodes share a high level of virtualisation. 
Whereas fewer cloud nodes represent fewer and much larger global resources at very long distances away from IoT devices, Fog nodes are numerous, smaller and more compact and importantly accessible from from IoT devices through local area or regional networks.  
IoT applications can be handled taking into account their specific requirements, such that low latency response, mobility support, location awareness, etc.

\emph{Contributions:}
We propose a framework for testing each IoT layer in a separate Virtual Machine (VM) environment. 
As per Bourque and Dupuis \cite{abran2004software}, software testing consists of the dynamic verification of the software behaviour against the expected behaviour, using a finite set of test cases. The set of test cases has to be selected from the (usually infinite) execution domain, with the goal to fulfil the specified coverage criteria. However, dynamic verification of program behaviour requires observation of its runtime behaviour (i.e., monitoring) thereby allowing measurement of compliance against expected behaviour of the system. 
Our study is focused on deriving operational profiles from monitoring data that is collected during dynamic verification of behaviour. We aim  (1) to use operational profiles for prioritising tests during test selection in regression testing, (2) to predict fault location in IoT services by combining operational profiles and Markov chain usage models, derived from interface behaviours of IoT services. 
If it is predicted that an IoT might have faults, it would undergo extended runtime verification procedure, accomplished with monitoring. 
The corresponding specifications for monitors are going to be derived from interface specifications of the standardised IoT services. 
We call this methodology ``monitoring informed testing''.

\section{IoT Architecture}
\label{sec:background}

IoT middleware could be divided into three layers (perception, middleware and application layers), where each of them has specific functionalities, cf. 
\cite{razzaque2016middleware}.

\emph{Perception Layer:} 
Sensor data i.e. context information is gathered from smart physical objects.  
IoT applications with various application-specific requirements 
(such as low latency response, mobility support, location awareness, etc.) are handled at this layer. 
Data is collected at collector nodes from sensors, then  filtered, processed, analysed, and after that the decided actions are performed through actuators. Functionalities like data acquisition, context annotation, device configuration, device management, etc., are provided as services through SOA-based IoT middleware. 

\emph{Middleware Layer:} Fog computing is considered as widely accepted approach that could be set up at this layer to prevent large burst of sensor data from flooding into the Internet. 
Fog nodes \cite{li2015internet} are highly virtualised platforms (e.g., IOx and Cloudlet), which may be a computer like ``cloud in a box'' providing computation, storage, and networking services. 
Middleware support  aims to provide higher-level services. 
For example, SOA IoT middleware provide activity reasoning, processing, machine-to-machine-interoperability services \cite{perera2014context}. 

\emph{Application Layer:} 
Global coverage for an IoT application can be provided through a Cloud. 
The Cloud is particularly useful to serve as data storage and also providing processing power needed for IoT big data analytics. 
Service interfaces, APIs, etc. provided through IoT middleware are utilised to implement data analytics, storage management, application and health monitoring, profiling, etc.     
Here, we explain the testing set-up of an industrial IoT application and also discuss SOA functionalities of industrial IoT middleware \emph{oneM2M} \cite{alaya2014om2m}, which is an %
industrial IoT middleware for machine-to-machine (M2M) interoperability, compliant to the European Telecommunications Standards Institute (ETSI) standards\footnote{\url{http://www.etsi.org/about}}.

\begin{figure}[ht!] 
\begin{center}
\centering
\includegraphics[scale=0.24]{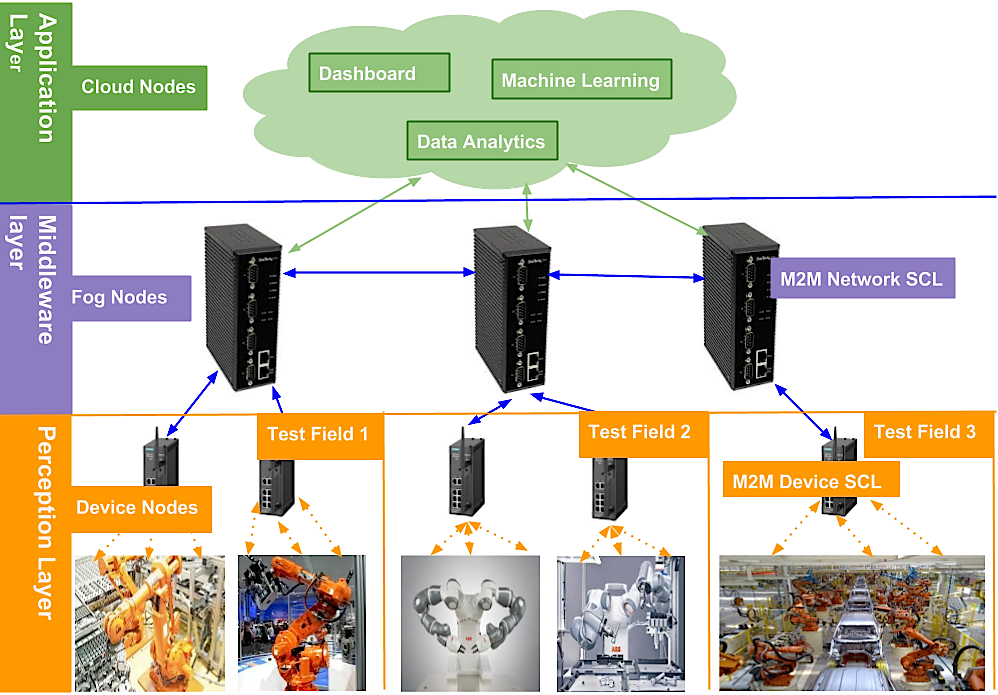}
\end{center}
\caption{Industrial IoT Networking Architecture}
\label{fig:fig1}
\end{figure}

\emph{Device Nodes:} 
We envision of a global industrial automation systems development company (e.g., ABB and Siemens) who has several \emph{Test Fields} in different countries across the world. 
In each \emph{Test Field} they would have a collection of IoT enabled smart devices for testing purposes. Such a \emph{Test Field} would have a variety of sensors and actuators attached smart machines, embedded devices and computational nodes. 
Figure~\ref{fig:fig1} illustrates perception layer setup for an industrial IoT application, 
where each smart machine is connected with a dedicated Device Service Capability Layer (DSCL), i.e., a middleware component provided by oneM2M that collects data from a smart device.

\emph{Fog Nodes:} 
We anticipate that \emph{Devices Nodes} that are located at a \emph{Test Field} will directly connect to the \emph{Fog Nodes}, cf. Figure \ref{fig:fig1}.  
OneM2M's Network Service Capability Layer (NSCL) is a middleware component that provides higher-level services such as resource access, provisioning, and self-configuration, etc.

\emph{Cloud Nodes:} The industrial IoT application would implement a dashboard providing data analytics, industrial process monitoring, machine learning, etc. at the Cloud. In oneM2M, nodes consist of at least one common services entity (CSE) or one application entity (AE). An AE defines application logic for an end-to-end M2M solution \cite{palattella2016internet}.

\section{Monitoring approaches}
\label{sec:monitoring}

The monitoring requirement for testing lies in the amount and type of information required for %
 (1) detecting failures due to bugs (functional or extra-functional violations) or slugs (performance bottlenecks),  
 (2) locating their root causes or errors,  and 
 (3) analysing the error context so as to prepare fixing or preventing them. 
For testing sequential processes \cite{thane2000monitoring}, it is usually sufficient
to monitor input and output behaviour through predefined service interfaces. 
The actual information to be monitored can be categorized into three groups: 
\begin{itemize}
\item \emph{Data flow} to observe the input data flow of a service as well as 
the output data  produced and passed through the service interfaces; 
\item 
\textit{Control flow} to observe in what order inputs are received and outputs are produced; and
\item
\emph{Resource flow} including usage and performance of memory, compute,  network and file system resources. 
\end{itemize} 
The services need to be instrumented with \emph{probes} to be able to observe their behaviours. 
Furthermore, these probes need to be left within the system. 
However, effect of probes in the system need to be measured and compensated by allocating additional resources to the system. 
However, \textit{in-line-probes},  a kind of probe instrumented at task level (e.g. atomic service level) to yield additional outputs to the task, would be sufficient for our purposes. Our goal is to create operational profiles through analysing the outputs that are produced through the service level in-line-probes instrumented within the system.
This approach has to be applied within the development phase. 
One of the potential issues related with the in-line-probes, is over instrumenting   that decreases the system performance as this may impact on Resource Flow monitoring. 

Another approach to monitoring is \emph{runtime verification} (RV), 
where a formal specification of behaviour is instrumented in the system during runtime. RV aims to combine testing with formal methods such that systems behaviour is thoroughly checked against requirement specifications at runtime. 
In this approach, formal models (e.g., state machines) for service interfaces are typically developed during designing test models. 
The formal model is derived from requirement specification and then integrated with source code. Afterwards, monitors are generated automatically based on the formal models and weaved into the runtime code. 
An advantage of RV is that formal models can be encoded with varying levels of abstraction thus allowing varying levels of monitoring data to be generated. 
%
\section{Statistical Testing}
\label{sec:testing}

In statistical testing, statistical methods are used to determine the reliability of a system  to demonstrate a system's fitness against its intended use. 
In our future work, we are going to focus on a fundamental approach that applies Markov Chain Usage Model (MCUM) in conjunction with this method. 
The statistical testing process may be started at any point: 
all test artefacts become valuable assets and may be reused throughout the software system life cycle.
Statistical testing involves several steps, such as usage model development, model analysis and validation, tool chain development, test case generation, etc. 
We are going to focus on the issues on usage model construction as it relates to one of the fundamental concepts of our study.

\subsection{Markov Chain Usage Model} 
A fundamental statistical artefact for statistical testing is study of population, which requires the characterization and representation of the system. This should include common and typical as well as infrequent and exceptional scenarios of the system at a suitable level of abstraction. One such method of characterisation and representation of system is done through developing operational usage model. A operational usage model characterises the population of usage scenarios described in terms of how it is going to be used in an operational environment. 
Markov Chain Usage Model (MCUM) characterises usually infinite population, which contains all possible scenarios of a system. However, when a population is too large for exhaustive study, a statistically correct sample must be drawn as a basis for inferences about the population.

MCUM is usually derived from requirements specification, and  
 can be developed in two stages, structural and statistical. The structural stage involves possible use; the statistical stage deals with expected use. Developing the structural model involves identifying sets of states within a system and associated state transitions, which are defined by directed arcs. The statistical stage involves determining transition probabilities in the structure. There are two basic ways of assigning transition probability - one based on direct assignment of probabilities and the other is determining values through analytical method.					

Direct assignment of transition probabilities among states in a usage model may involve collecting data from historical or projected usage of an application. Transition probabilities among states yield various usage information of a system for example usage environments, user demographics, classes, or other special usage situations. 
Moreover, sets of transition probability may change several times as systems mature, based on availability of information, or experience of use. 
A probability value for each arc of the model may be determined when extensive field data for systems is collected over long periods of time. 
However, to determine transition probabilities for new systems, we may follow a manual process, 
i.e., analyse software requirements documents, review user guides, 
take customer reviews, etc. 
In case essential information is not available, all transition states are assigned with uniform probability in the beginning. 

\subsection{Operational profile} 
An operational profile characterizes how a system will be used in production \cite{musa1993operational}. Therefore, it allows estimating reliability of a software product, helps prioritization of product feature to be developed and tested 
according to their usage frequency. In the context of IoT systems and software in production, deployed across millions of devices with thousands of different variants of code and personal user preference settings, it is particularly important to reflect the frequency of software deployment in specific device, location and personal contexts. This does not only require monitoring the unconditional probability of executing a piece of code globally, but its conditional probability, given the likelihood of executing in a specific context defined by variants of devices, OS, application software, location and personal preferences etc.

When software is developed, in testing phase or already in deployment, operational profiles may be generated either manually as described earlier or automatically through analysing event logs. The operational profile then may be used to generate random test cases such that random walk through MCUM \cite{poore2011automated}.

\subsection{Continuous Integration and Regression Testing} 
An IoT application may be layered into the three  distinguishable components: 
\emph{Device Component} in perception layer, 
\emph{Middleware Component} in networking layer, and 
\emph{Application Component} in presentation layer. Each of these functional components would undergo automated regression testing as soon as new changes to the code are submitted. %
Regression testing nowadays is integrated with continuous integration software development practice that aims to ensure program's correctness as soon as new changes to the code is submitted to a mainstream code repository that triggers automated build, and testing. 
One of the core functionalities of regression testing is prioritizing test cases to achieve some performance goal~\cite{zhang2014application}, for example selecting frequently used features for testing early. Executing tests ordered by priority of frequently used feature would allow to achieve higher reliability by discovering and fixing high frequency failures sooner. Hence, we aim to prioritize tests according to the order of frequency of the features derived within an operational profile.

\section{Framework for Monitoring Informed Testing} 
\label{sec:framework}

The aim of the framework is to monitor IoT services and generate operational profile from monitoring information.  IoT systems provide various types of services at different layers. Sensor data collected through service interfaces travel from perception layer, through middleware layer to application layer. Furthermore, the data is formatted and processed at different layers differently. 
Our goal is to analyse the monitoring information in services as well as in sensor data with a view to design operational profile for an IoT services. We aim to design continuously adapted operational profiles. Thus, we need to investigate in what time intervals the operational profiles are adapted and also that the same event is not recorded multiple times in the contentiously adapted profile.

We are going to prioritize tests based on service usage frequency in operational profile. 
The monitors will be derived from specifications of IoT services and integrated into relevant IoT services at IoT runtime. Our intention is to apply runtime verification only for the services identified through fault prediction mechanism. 
Thus, MCUM will be derived the interface specifications that are provided with IoT services for identifying states and state-transitions that are needed for a MCUM.   
The judgements from operational profiles have to be evaluated against the prediction derived through MCUM. Therefore, we plan to design a high confidence fault prediction mechanism where prediction measured through MCUM would be combined with that gathered from the \emph{operational profile}.         

\begin{figure}[ht!]
\begin{center}
\centering
\includegraphics[scale=0.24]{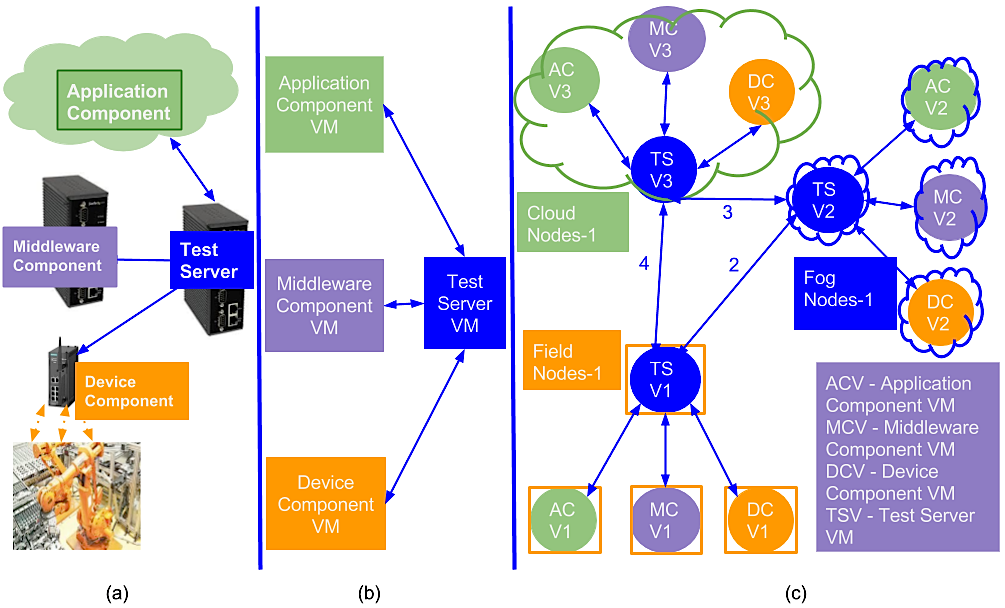}
\end{center}
\caption{Proposed Framework: Infrastructure}
\label{fig:fig2}
\end{figure}

All three IoT architectural layers (i.e., perception, middleware and application layers) 
will be tested in separate VMs, as illustrated in Figure \ref{fig:fig2}.
We define a \emph{Test Server}  VM required for Continuous Integration (CI) setup,  involving loading source code and tests from repository, doing the builds, deploying the build to appropriate test machines and running regression tests. 

A global company with multiple test fields would require a large test network setup with test VMs deployed across Fog and Cloud. 
Figure \ref{fig:fig2} shows a network of test VMs that forms a weighted graph where weight of an edge is measured according QoS constraints like bandwidth, latency, etc. Thus, we intend design an optimal VM allocation strategy that would allocate VMs in test network considering all these QoS constraints. Such an VM allocation method is needed for various reasons such as lowering the amount of data transfer through high cost edge, thereby enabling faster test execution, lowering networks usage, etc. However, the VM allocation method needs to consider following issues:
\begin{itemize}
\item
\emph{Data exchange:} 
A huge amount of data would be exchanged across the test network. 
For example source code and tests need to be transferred to test servers, 
and tests have to be sent on testing VMs for execution. 
An optimal VM allocation method would be required to minimise the amount of data transferred, especially across the high cost edge.
\item
\emph{Allocation of VMs:} 
Device Components (DCs) have to be tested at the field network at perception layer.  
Virtual sensors may be suitable for testing purposes to simulate data into DCs. 
This would allow us to allocate them either to the Fog or the Cloud.  
In some cases where testing of virtual sensors or actuators is not feasible, 
the VM allocation method should find an optimal route keeping DC nodes allocated to a dedicated layer.
\item
\emph{Quality of Service:} 
Network Quality of Service (QoS) constraints such as bandwidth, latency and network failure have to be considered. 
For example, the network connection between \emph{Field Node} and \emph{Fog Node} should have high bandwidth, low latency, and very low network failures. 
Therefore, the QoS constraints need to be measured as part of our Resource Flow monitoring and as demanded by the optimization method.  
\end{itemize}

\subsection{Operational profile based online test prioritization}

Software testing based on users' perspective is known as usage-based testing. 
Both Markov chains and operational profiles are used to characterise usage models statistically. 
Many usage based testing methods have been proposed in classical software engineering and also a few in the service oriented architectures. 
Sammodi et al.\cite{sammodi2011usage} applied operational profile for test case prioritization where service based monitoring data is used to derive operational profile. 
From an IoT perspective, data is generated from heterogeneous smart devices and collected through IoT services.  Hence, we believe it is essential to analyse monitoring data with reference to appropriate inline probing mechanism that we discussed previously. 
Bai et al.\cite{bai2008ontology} proposed an ontology-based method for producing operational profile through analysing SOAP messages. However, they use the operational profile for test generation purposes. On the other hand, we aim to create operational profiles for IoT oriented CoAP or REST services for the purposes of test prioritization. 

There are also many approaches on operational profile generation. 
In the industry, various tools are used for event correlation, e.g., SEC \cite{vaarandi2002sec}.  
Nagappan et al.\cite{nagappan2009efficiently} proposed an algorithm for creating operational profiles that calculates frequency of both used and unused features. The procedure begins with identifying the list of functions from the source code that prints out logs. Then a log abstraction method is used, where an integer equivalent of the execution log is created. Afterwards, using that integer equivalent of the execution log an operational profile is generated for each identified function by implementing a suffix array based algorithm that has \emph{log(N)} complexity. However, our goal is to produce continuously adapted operational profile from monitoring data and event logs generated from IoT devices.

\subsection{Usage model based fault prediction}

The main objective of statistical testing methods is reliability prediction. Usage models such as Markov chains and operational profiles are built to achieve this objective. Test cases generated from usage modes carry statistical significance and failure of test case (test sequence) is referenced back to usage model for predicting the location of fault. Sammodi et al.\cite{sammodi2011usage} cross checked test sequence from Markov chain with input sequence in monitoring information and based on some precision measurement in fault prediction and propose runtime adaptation of services. 
Metzger et al.  \cite{metzger2010towards} applied similar fault prediction mechanism, also suggesting runtime testing of services for confidence measurement. In contrast to these approaches, our goal is to enable runtime verification of service that is predicted to have fault. 


\subsection{Virtualized IoT layers and Optimal VM Allocation}

Integrating heterogeneous smart objects (e.g. sensors and actuators) to an automated test system would be difficult task. Also setting up a test system and maintaining it with connected smart devices would be challenging. We aim to use virtual sensors \cite{reetz2016service} to simulate the interactions of smart objects with the physical layer of IoT middleware, thereby enabling the Physical Component of IoT application be tested in a VM. 
A test network for a global industrial IoT company could be depicted as a weighted graph, cf. Figure \ref{fig:fig2}. 
A similar scenario where optimal VM allocation problem over distributed Cloud been presented as a weighted graph in \cite{alicherry2012network}. The authors explained this to be a graph slicing (a NP-hard) problem and proposed two approximation algorithms, one for VM allocation within same cloud and another for inter cloud VM allocation.  However, we are considering a weighted graph which has multiple dynamic weights, i.e., bandwidth, latency, network failure, etc. and also the weight of an edge would be calculated dynamically.

\section{Conclusions}
\label{sec:conclusions}

This paper introduces the core ideas of an adaptive framework for monitoring informed testing, with an aim to monitor IoT services and generate operational profile from monitoring information. 
The proposed framework will enhance testing activities by  utilizing monitoring data gathered from IoT smart devices and relevant event logs. 
We suggest 
\begin{itemize}
\item[(1)] to use monitoring probes for Data, Control and Resource flow combined;
\item[(2)] to model operational profiles in a context dependent way using conditional  probabilities with reflecting the high variation in hardware, software and consumer contexts, and 
\item[(3)] to predict fault locations  in IoT services by combining operational profiles and Markov chain usage models, derived from interface behaviours of IoT services. 
\end{itemize} 
This would 
enable, e.g., runtime verification of service that is predicted to have fault.
 
\balance

\bibliographystyle{IEEEtran}

\end{document}